\documentstyle[a4wide,11pt]{article}
\title{Quantum mechanics in multiply connected spaces}
\author{Vu B Ho and Michael J Morgan\\Department of Physics\\
Monash University\\Clayton Victoria 3168\\Australia}
\date{}
\begin{document}
\maketitle
\begin{abstract}
This paper analyses quantum mechanics in multiply connected spaces. It is
shown that the multiple connectedness of the configuration space of a
physical system can determine the quantum nature of physical observables,
such as the angular momentum. In particular, quantum mechanics in
compactified Kaluza-Klein spaces is examined. These compactified spaces
give rise to an additional angular momentum which can adopt half-integer
values and, therefore, may be identified with the intrinsic spin of a
quantum particle.
\end{abstract}

\newpage

\section{Introduction}

Quantum mechanics conventionally deals with the evolution
of a particle in a simply connected configuration spaces, whose topology is
Euclidean $\mbox{R}^n$. The Euclidean space $\mbox{R}^n$ has a simple
topology in the sense that paths in this space are contractible to a point,
and so the fundamental homotopy group is trivial, i.e. $\pi_1(\mbox{R}^n)
\cong \{0\}$. However, novel features can arise from configuration spaces
that exhibit a nontrivial topology \cite{Isham,Avis}.
In a multiply connected space, the nature of the Schr\"{o}dinger wavefunction 
may depend on the topological structure of the configuration space.
On the other hand, multiply connected configurational
spaces have also been shown to play an important role in field theory and
particle physics, where stable solutions to field equations can exist for
appropriate topologies of the configuration spaces. \cite{Bala,Ryde}.

In this paper we analyse quantum mechanics in multiply connected
configuration spaces. It is assumed that a physical system is described by a
complex wavefunction $\Psi$ defined on a configuration space $M$, and the
time evolution of the system is determined by the Schr\"{o}dinger equation
\begin{equation}
i\hbar\frac{\partial \Psi}{\partial t} = H\Psi,
\end{equation}
where $H$ is the Hamiltonian of the system. According to the orthodox
interpretation of the wavefunction in quantum mechanics, all of the
physical information is contained in the square of the modulus of the
wavefunction, therefore, all complex wavefunctions that differ from one
another by a phase can be used to describe the same quantum state of the
physical system. Consequently, the absolute phase is regarded as
unobservable and the phase of a wavefunction can be defined globally as a
U(1)-valued function on the configuration space $M$, when $M$ is simply
connected. In terms of a fibre bundle structure, this result can be seen as
an admission of a global section of a principal fibre bundle, since
nonrelativistic quantum mechanics can be associated with the trivial complex
line bundle $M\times C$ and the wavefunction is simply a global section of
the line bundle. The quantum state of a physical system is regarded as an
equivalence class $\{e^{i\alpha}\psi\}$ of normalised wave functions
defined on the principal U(1)-bundle $M\times S^1$ on $M$. Because
physical observables correspond only to the squared modulus of a wave
function, wave functions on the U(1)-bundle must satisfy the condition
$\psi({\bf x},e^{i(\alpha+\beta)})=\psi({\bf x},e^{i\alpha})e^{i\beta}$
\cite{Bala,Mora,Naka}.

When the topological structure of the configuration space
is nontrivial, ambiguities may arise when attempts are made
to specify a value for the phase of a wavefunction for the whole
configuration space. In this case the fibre bundle representation of
the system is not trivial and so it does not admit a global section.
However, since physical observables related to $\psi^*\psi$ are considered
as functions on the configuration space, wavefunctions which are functions
on a fibre bundle over the configuration space are allowed, provided they
satisfy the requirement $\psi({\bf x}e^{i\alpha})=\psi({\bf
x})e^{in\alpha}$. This problem is related to the problem of formulating 
quantum mechanics on the universal covering space of a multiply connected
configuration space. This paper is organised as follows. In Section 2
a brief review of quantum mechanics on universal covering spaces is given.
Section 3 discusses the quantisation of angular momentum. Section 4
discusses quantum mechanics on a multiply connected spaces. In particular
we show that quantum mechanics in compactified Kaluza-Klein spaces 
gives rise to an additional angular momentum which adopts 
half-integer values, and therefore may be identified with the intrinsic
spin of a particle.

\section{Quantum mechanics on universal covering spaces}

Consider the case where the configuration space $M$ of a physical system is
a topological space which may be multiply connected, so that its
fundamental homotopy group $\pi_1(M)\neq\{0\}$. If the space $M$ satisfies
the conditions of arcwise and local connectedness, then it is possible to
construct a covering space $\widetilde{M}$ which is simply connected with
the property $\pi_1(\widetilde{M})=\{0\}$. The covering space
$\widetilde{M}$ is called a universal covering space of $M$.
The space $\widetilde{M}$ is a bundle space over the space $M$ with a
covering projection $\pi:\widetilde{M}\rightarrow M$ so that the homotopy
group of the bundle space $\widetilde{M}$ is a factor group of the
fundamental homotopy group of the base space $M$ \cite{Steen,Sing,Hilton}.
Although it is always possible to construct a conventional quantum
mechanics on the universal covering space $\widetilde{M}$, because it is
simply connected, the question of how to relate the quantum mechanics on
$\widetilde{M}$ to quantum mechanics on the multiply connected space $M$
does not have an obvious answer. Consider a single-valued wavefunction
$\widetilde{\psi}(\widetilde{\bf x})$ on $\widetilde{M}$. If the point
${\bf x}=\pi(\widetilde{\bf x})$ on $M$ is taken around a loop $\gamma$,
then when the loop $\gamma$ lifts to a curve $\widetilde{\gamma}$ in the
space $\widetilde{M}$, with the initial point $\widetilde{\bf x}$, the wave
function $\widetilde{\psi}$ will take its value at the end point
$\widetilde{\bf x}'$ of the curve $\widetilde{\gamma}$, obtained from the
point $\widetilde{\bf x}$ by the action of the homotopy class $[\gamma]$,
that is, $\widetilde{\psi}(\widetilde{\bf
x}')=\widetilde{\psi}([\gamma]\widetilde{\bf x})$. Therefore, if
quantum mechanics on the bundle space $\widetilde{M}$ is projectable to 
quantum mechanics on the multiply connected space $M$, in the sense that
the squared modulus of the wave function $\widetilde{\psi}$ on
$\widetilde{M}$ depends only on ${\bf x}=\pi(\widetilde{\bf x})$, then the
wavefunction $\widetilde{\psi}$ must satisfy the boundary conditions
\begin{eqnarray}
\widetilde{\psi}([\gamma]\widetilde{\bf x}) &=& a([\gamma])\widetilde{\psi}
(\widetilde{\bf x})\\
|a([\gamma])|&=&1
\end{eqnarray}
for all $\widetilde{\bf x}\in\widetilde{M}$ and all homotopy classes
$[\gamma]\in \pi_1(M)$. Furthermore, because the wavefunction
$\widetilde{\psi}$, being defined on a simply connected space, is
single-valued, the phase $a([\gamma])$ must also satisfy the condition
$a([\gamma])a([\gamma'])=a([\gamma][\gamma'])$. The conditions imposed on
the phase $a([\gamma])$ show that the map $a:\pi_1(M)\rightarrow U(1)$
defined by $[\gamma]\rightarrow a([\gamma])$ is a one-dimensional unitary
representation of the fundamental homotopy group $\pi_1(M)$ \cite{Mora}.

To illustrate how quantum mechanics on a multiply connected space can be
realised in physics, let us consider the quantum dynamics of a particle
moving in a one-dimensional lattice with a periodic potential
$V(x+nd)=V(x)$. Assume the dynamics is governed by a Hamiltonian of the
form $H=p^2/2m+V(x)$, where $m$ and $p$ are the mass and the momentum of
the particle, respectively. The system in this case has translational
symmetry, since the Hamiltonian is invariant under the transformation
$x\rightarrow x+nd$. Let $T(n)$ be an operator that corresponds to this
transformation in the vector space of physical states of the system, then
it can be shown that the set $\{t_n(\theta)=e^{-in\theta};
~-\pi\leq\theta<\pi\}$, forms a one-dimensional unitary representation of
the translation group $T(n)$. It follows that the Schr\"{o}dinger
wavefunctions of the particle, known as Bloch functions, must satisfy the
boundary condition $\psi_{nk}(x+d)=\exp(ikd)\psi_{nk}(x)$. Now if the
endpoints of a unit cell of the lattice are identified, so that the cell
has the topology of a circle $S^1$, which has the  fundamental homotopy
group isomorphic to Z, then it is seen that the characters are
$a([\gamma])=\exp(ikd),~-\pi\leq kd<\pi$. Another illustration of quantum
mechanics on multiply connected space is that of the hydrogen atom in the
Euclidean plane R${}^2$. The time-independent wavefunction for the
system is written in the form $\psi(r,\phi)=\exp(im\phi)R(r)$, where $R(r)$
is the radial solution. When the configuration space of the system is
simply connected, the wave function $\psi$ must be single-valued, which
imposes the requirement that the quantity $m$ must be integer. However, if
the electron of the system can not penetrate the nucleus, it is reasonable
to reduce the space to R${}^2\setminus\{0\}$, which is now multiply
connected, and whose fundamental homotopy
group is isomorphic to $\pi_1(S^1)$. Therefore, the wavefunction on the
original space R${}^2$ is projectable even if the quantity $m$ takes
half-integer values; in this case the wavefunction satisfies the
boundary condition $\psi(r,[\gamma]\phi)=\psi(r,n\phi)=\exp(in\pi)
\exp(im\phi)R(r)$ and hence $a([\gamma])=a(n)=\exp(in\pi)$. This problem
will be discussed in more detail in the next section.

Having shown how to construct a projectable quantum mechanics on a
universal covering space $\widetilde{M}$ of a multiply connected
configuration space $M$, there still remains the question: What is the 
nature of
the wavefunctions on $M$? The universal covering space $\widetilde{M}$
is the union of fundamental domains, each of which is isomorphic to the
configuration space $M$. If quantum mechanics on the universal covering
space $\widetilde{M}$ is restricted to that of a particular domain, then
the wavefunction $\widetilde{\psi}$ can be projected down to a well-defined
wavefunction $\psi$ on the space $M$. However, since a point ${\bf x}\in M$
corresponds to many
different points $\widetilde{\bf x}=\pi^{-1}({\bf x})\in \widetilde{M}$,
which are connected by the action of the fundamental homotopy group
$\pi_1(M)$ on $\widetilde{M}$, the projection of the wave function
$\widetilde{\psi}$ on $\widetilde{M}$ to a wave function $\psi$ on $M$
will make the wave function $\psi$ multivalued. The construction of a
projectable quantum mechanics on universal covering spaces requires that the
multiple valuedness of a wavefunction on the original configuration space
(obtained by the projection of a wave function on a universal covering
space) is not arbitrary, but is limited to multiplication
by the characters of the fundamental homotopy group of the original
configuration space \cite{Bala,Mora}.

\section{On the quantisation of angular momentum}

As mentioned earlier the quantum mechanics of a hydrogen atom in the
configuration space $\mbox{R}^2\setminus \{0\}$ can be considered in terms
of a multiply connected space whose first fundamental homotopy group is
isomorphic to $\pi_1(S^1)$. In this case, wavefunctions on the space
$\mbox{R}^2\setminus \{0\}$ are allowed to be multivalued \cite{Ho}. Here
we examine the relationship between the multiple valuedness of the
wavefunction and the quantisation of angular momentum. Consider the
eigenvalue equation of a generalised hydrogen-like atom consisting of a
single electron of charge $-e$ and a nucleus of charge $Ze$
\cite{Hagen,Louck,Nieto}
\begin{equation}
-\frac{\hbar^2}{2\mu}\nabla_N^2\psi({\bf r}) - \frac{Ze^2}{r}\psi({\bf r}) =
E\psi({\bf r}),
\end{equation}
where $\mu$ is the reduced mass, $r=\sqrt{g_{\mu\nu}x^\mu x^\nu}$ and
$\nabla_N^2=g^{\mu\nu}\partial_\mu\partial_\nu$, where $g_{\mu\nu}$ is the
$N$-dimensional Euclidean metric. In an $N$-dimensional Euclidean space, 
spherical polar coordinates are defined in terms of the coordinates
$\theta_i$, $1\leq i\leq N-1$, i.e.
\begin{eqnarray}
x_1&=&r\cos\theta_1\prod_{i=2}^{N-1}\sin\theta_i \nonumber\\
x_2&=&r\prod_{i=1}^{N-1}\sin\theta_i \\
x_i&=&r\cos\theta_{i-1}\prod_{j=i}^{N-1}\sin\theta_j \ \ \ \ i=3,...,N
\nonumber
\end{eqnarray}
where the range of the variables is $0\leq r<\infty$, $0\leq \theta_1\leq
2\pi$ and $0\leq \theta_i\leq\pi$ for $i=2,...,N-1$. It is straightforward
to verify that the Laplacian $\nabla_N^2$ expressed in terms of the $N$-
dimensional polar coordinates takes the form
\begin{equation}
\nabla_N^2=\frac{1}{r^{N-1}}\frac{\partial}{\partial r}r^{N-1}
\frac{\partial}{\partial r} - \frac{L_{N-1}^2}{\hbar^2r^2},
\end{equation}
where the generalised angular momentum operators are defined by the
recursion relation
\begin{equation}
L_{i-1}^2=-\hbar^2\left\{\frac{1}{\sin^{i-2}\theta_{i-1}} \frac{\partial}
{\partial\theta_{i-1}}\sin^{i-2}\theta_{i-1}\frac{\partial}{\partial
\theta_{i-1}} - \frac{L_{i-2}^2}{\hbar^2\sin^2\theta_{i-1}}\right\}.
\end{equation}
The generalised spherical harmonics $Y(l_{N-1},...,l_1)$ are defined as the
simultaneous eigenfunctions of the set of operators $\{L_k^2\}$
\begin{equation}
L_k^2Y(l_{N-1},...,l_1)=l_k(l_k+k-1)\hbar^2Y(l_{N-1},...,l_1),
\end{equation}
where $1\leq k\leq{N-1}$. Imposing the requirement of single-valuedness,
the quantum number $l_i$ with $1\leq i\leq N-1$ must be an integer.
On the other hand, integer values of the quantum number $l$ are required for
consistency of the group representation \cite{Merz,Land}. When the
eigenfunction of the generalised hydrogen atom is written as
\begin{equation}
\psi_{nl}(l,l_{N-2},...,l_1) = R_{nl}Y(l,l_{N-2},...,l_1),
\end{equation}
with $l=l_{N-1}$, then the radial equation for the function $R_{nl}$, for the
case of bound states with $E<0$, becomes
\begin{equation}
\frac{d^2R_{nl}}{d\rho^2}+\frac{N-1}{\rho}\frac{dR_{nl}}{d\rho} -\left[
\frac{l(l+N-2)}{\rho^2}+\frac{\lambda}{\rho}-\frac{1}{4}\right]R_{nl} = 0,
\end{equation}
where $\rho$ and $\lambda$ are defined by
\begin{equation}
\rho=\left[\frac{8\mu(-E)}{\hbar^2}\right]^{1/2}r, \ \ \ \ \
\lambda=\left[\frac{Z^2e^4\mu}{2\hbar^2(-E)}\right]^{1/2}.
\end{equation}
We seek solutions for $R_{nl}$ of the form
\begin{equation}
R_{nl}=\exp(-\rho/2)\rho^lS(\rho).
\end{equation}
By substitution into the equation for $R_{nl}$ the following
differential equation for $S(\rho)$ is obtained, i.e.
\begin{equation}
\frac{d^2S}{d\rho^2} + \left(\frac{2l+N-1}{\rho}-1\right)\frac{dS}{d\rho} +
\frac{\lambda-(N-1)/2-l}{\rho}S=0.
\end{equation}
This equation can be solved by a series expansion of $S(\rho)$
\begin{equation}
S(\rho)=\sum_{n=0}^\infty a_n\rho^n,
\end{equation}
with the coefficients $a_n$ satisfying the recursion relation
\begin{equation}
a_{n+1}=\frac{n+l+(N-1)/2-\lambda}{(n+1)(n+2l+N-1)}a_n.
\end{equation}
The bound state energey spectrum is given by
\begin{equation}
E_n=-\frac{Ze^4\mu}{2\hbar^2}\frac{1}{[n+l+(N-1)/2]^2}.
\end{equation}
This result shows that for spaces of odd dimension the quantum number
$l$ must be an integer for the energy $E_n$ to have the same form as that of
the Bohr model. On the other hand, for spaces of even dimension, the Bohr
spectrum is obtained only when the quantum number $l$ is half-integer.
However, in quantum mechanics the quantum
number $l$ must always be integer by the requirement of single-valuedness
of the Schr\"{o}dinger wave function, irrespective of the dimension of the
configuration space. We now
discuss this problem from the perspective of quantum mechanics on
multiply connected spaces, showing the important relationship between the
topology of the system and the multiply connected configuration
space of the atom. In two-dimensional space, the Schr\"{o}dinger equation
in planar polar coordinates takes the form
\begin{equation}
-\frac{\hbar^2}{2\mu}\left[\frac{1}{r}\frac{\partial}{\partial r} \left(r
\frac{\partial}{\partial r}\right) + \frac{1}{r^2}
\frac{\partial^2}{\partial \phi^2}\right]\psi(r,\phi)-
\frac{Ze^2}{r}\psi(r,\phi) = E\psi(r,\phi),
\end{equation}
where it is assumed that the Coulomb potential has the $r^{-1}$ form. This
equation admits solutions of the form $\psi(r,\phi)=R(r)\exp(im\phi)$, 
where $m$ is identified with the angular momentum of the system.
We normally require the angular momentum $m$ to take integer values so that
the single-valuedness condition is satisfied. However, the requirement
that $m$ be integer is not compatible with the assumption that an observer
in a two dimensional space must obtain an energy spectrum identical to the
Bohr model because the energy spectrum in this case can be written
explicitly in the form
\begin{equation}
E=-\frac{Z^2e^4\mu}{2\hbar^2(n+m+\frac{1}{2})^2}.
\end{equation}
Hence, if the hydrogen-like atom is viewed as a two dimensional physical
system, and if the energy is observed to have the same spectrum as that of
the Bohr model then the angular momentum $m$ must take half-integer
values. These half-integer values are allowable provided the configuration
space is not simply connected. This will be the case for a hydrogen-like
atom viewed as a planar system, in which the atomic electron cannot
penetrate the nucleus. In this situation the single-valuedness condition is
no longer a sufficient requirement to ensure that the angular momentum 
adopts integer values. However, it can be verified that integer values for
the angular momentum $m$ can be retained if we add to the Coulomb potential
a quantity $-\left(\hbar\sqrt{E/2\mu}\right)/r$, when the hydrogen-like atom
is viewed as a two dimensional physical system.

In the above example it is seen that topological structure of a
configuration space can determine the quantum nature of an observable. This
result is not unexpected in quantum mechanics. If the electron in a
hydrogen-like atom is constrained to move in a plane, then the orbital
angular momentum of the electron must take half-integer values in order to
reproduce the same energy spectrum as the Bohr model. As a consequence, it
may be possible to invoke topological constraints to explain the
Stern-Gerlach experiment, without the necessity of introducing spin into
non-relativistic quantum theory in an {\it ad~hoc} manner. In the following
sections we address these issues in more detail.

\section{Quantum mechanics in compactified Kaluza-Klein spaces}

The topological structure of a physical system can determine the nature of
an observable of a quantum system, such as angular momentum.
In the case of a hydrogen-like atom, the nontrivial topological structure of
the system can only be revealed when the electron is constrained to move in
a plane, so that the fundamental homotopy group,
$\pi_1(R^2\setminus\{0\})$, is nontrivial. This results in the 
angular momentum
adopting have half-integer values, since the wavefunction in this case is
allowed to be multi-valued. However, there remains the problem of how to
incorporate topological constraints into the dynamics of the electron in
three-dimensional space.  One possible approach is to use path integral
methods in multiply connected spaces, where spin can be incorporated
by specifying an appropriate space, e.g. SO(3). In this manner, continuous
classical mechanics, when defined and quantised, can provide a
framework for incorporating spin \cite{Schu,Schul}. However, in the present
work it is desired that the topological description should only involve
spacetime structures. Consider the quantum mechanics of a generalised
$N$-dimensional hydrogen atom whose bound state spectrum given by
$E_n=-\mu e^4/(2\hbar^2[n+(N-3)/2]^2)$. It is noted that for spaces of
even dimensions the Bohr energy spectrum is retained only if the angular
momentum adopts half-integer values. This energy spectrum is derived for a
hydrogen atom in the simply connected $N$-dimensional Euclidean space $R^N$
whose fundamental homotopy group is trivial, i.e. $\pi_1(R^N)\cong\{0\}$.
In the simply connected Euclidean space R${}^N$, the wavefunction must be
single-valued, and, as a consequence, the angular momentum must be integer.
However, for quantum mechanics in multiply connected spaces, the Euclidean
space R${}^N$ may be considered as a universal covering space of some
multiply connected space in which a wavefunction of the Schr\"{o}dinger
equation can be multivalued. It is known that the Euclidean space R${}^N$
is a universal covering space of the space R${}^{N-1}\times S^1$
\cite{Naka,Sing}. The space R${}^{N-1}\times S^1$ has a nontrivial 
topological structure because its fundamental homotopy group is isomorphic
to Z, i.e. $\pi_1(R^{N-1}\times S^1)\cong \pi_1(R^{N-1})\oplus \pi_1(S^1)
\cong \{0\}\oplus Z\cong Z$. The multiply connected space R${}^{N-1}\times
S^1$ has the structure of a Kaluza-Klein space, because, according to modern
perspectives, a Kaluza-Klein space is not considered as an $M^N$ manifold
whose symmetries are the $N$-dimensional Poincar\'{e} symmetries, but rather
a compactified manifold of the form $M^4\times S^d$. Here $M^4$ is
four-dimensional Minkowski spacetime and $S^d$ is some compact manifold
whose size is much smaller than any length that has ever been measured
\cite{Kalu,Mahe}. In the following sections we consider nonrelativistic
quantum mechanics in a compactified Kaluza-Klein space $R^{N-1}\times S^1$
consisting of the direct product of an $(N-1)$-dimensional Euclidean space
R${}^{N-1}$ with the compact circle $S^1$. We assume quantum mechanics is
valid in these compactified spaces. The introduction of the compact circle
makes it possible to incorporate spin into the nonrelativistic
Schr\"{o}dinger wave equation. Since the topological structure
of the configuration space of a physical system depends on the dimension of
the space of the system, we discuss two, three and four-dimensional
compactified spaces separately.

\subsection{Quantum mechanics in 2-dimensional space $R^1\times S^1$}

Let us consider first the case of quantum mechanics in a two-dimensional
Kaluza-Klein space $R^1\times S^1$, where the compact space $S^1$ is a
circle of radius $\rho$ and $R^1$ is a one-dimensional Euclidean space.
This space has the form of a cylinder of radius $\rho$ embedded in the
three-dimensional Euclidean space $R^3$. The time-independent
Schr\"{o}dinger wave equation for a free particle
moving in this space can be written as
\begin{equation}
-\frac{\hbar^2}{2\mu}\left(\frac{1}{\rho^2}\frac{\partial^2}{\partial\phi^2}
 + \frac{\partial^2}{\partial z^2}\right)\psi(z,\phi) = E\psi(z,\phi).
\end{equation}
When the wavefunction is written in the separable form
$\psi(z,\phi)=Z(z)\Phi(\phi)$, the above Schr\"{o}dinger equation is reduced
to the following system of differential equations
\begin{eqnarray}
\frac{d^2\Phi}{d\phi^2}+s^2\Phi&=&0,\\
\frac{\hbar^2}{2\mu}\frac{d^2Z}{dz^2} + \left(E - \frac{\hbar^2s^2}
{2\mu\rho^2}\right)Z &=& 0,
\end{eqnarray}
where the quantity $s$ may be identified with the angular momemtum of the
system. The solution for the function $\Phi$ is of the form
$\Phi(\phi)=\exp(is\phi)$. In
order to obtain nontrivial solutions, the energy of the particle must
satisfy the condition $E\geq s^2\hbar^2/2\mu\rho^2$. In this case, the
solution is of the form $Z(z)=\exp(ikz)$, where $k$ is
a real number defined via the relation
\begin{equation}
E = \left(k^2+\frac{s^2}{\rho^2}\right)\frac{\hbar^2}{2\mu}.
\end{equation}
It is interesting to note that free
particles in this compactified Kaluza-Klein space can posses an angular
momentum $s$ which can adopt half-integer values, since the space $R^1\times
S^1$ is multiply connected.
This result allows an interpretation of the spin of a particle as a
manifestation of the topological structure of spacetime at the quantum
level. It is also noted that the ground state energy
$E=s^2\hbar^2/2\mu\rho^2$ of a free particle in this space is very large if
the size of the compact space $S^1$ is very small. However, if the size of
the compact space is not measurable then this energy is unobservable
because it is associated only with the compact space.

Now consider a particle under the influence of a Coulomb-like potential
which depends only on distance. Let the nucleus of positive charge $e$ of
a hydrogen-like atom be at the origin of the space R${}^1\times S^1$;
the distance from the nucleus to the atomic electron
is $r=\sqrt{z^2+(\rho\phi)^2}$. In this case the Schr\"{o}dinger equation
takes the form
\begin{equation}
-\frac{\hbar^2}{2\mu}\left(\frac{1}{\rho^2}\frac{\partial^2}
{\partial \phi^2} +
\frac{\partial^2}{\partial z^2}\right)\psi - \frac{e^2}{r}\psi = E\psi.
\end{equation}
We assume that the size of the compact manifold is small so that the
condition $\rho\ll z$ can be imposed. This allows us to expand the Coulomb
potential and use perturbation theory to calculate the first order
correction to the energy spectrum. Consider the electron
confined to the region $z>0$, which is
equivalent to a potential of the form \cite{Nieto}
\begin{equation}
V(z)=\left\{\begin{array}{ll}
-\frac{e^2}{z}+\frac{e^2\rho^2\phi^2}{2r^3}-
\frac{3e^2\rho^4\phi^4}{8r^5}+... & \mbox{for $z>0$}\\
+\infty & \mbox{for $z\leq 0$}
\end{array} \right.
\end{equation}
The terms involving $\rho\phi$ are treated as a perturbation. The 
unperturbed Schr\"{o}dinger equation takes the form
\begin{eqnarray}
\frac{d^2\Phi}{d\phi^2}+s^2\Phi&=&0,\\
\frac{\hbar^2}{2\mu}\frac{d^2Z}{dz^2}+\left[\frac{e^2}{z}+\left(E -
\frac{\hbar^2s^2}{2\mu\rho^2}\right)\right]Z&=&0.
\end{eqnarray}
The solution to Eq.(25) is $\Phi(\phi)=\exp(is\phi)$. On the other hand,
Eq.(26) represents the time-independent Schr\"{o}dinger equation for a
one-dimensional hydrogen atom. The solution is  given by
\begin{equation}
\psi_n(z)=\left[\frac{(n-1)!}{2n(n!)^3}\right]^{1/2}e^{-u/2}uL_n^1(u)
\end{equation}
where $L_n^1(u)$ is the associated Laguerre polynomial and
$u=2\mu e^2z/\hbar^2n$.
The bound state energy spectrum in this case is given by
\begin{equation}
E_n=-\frac{\mu e^4}{2\hbar^2n^2}+\frac{\hbar^2s^2}{2\mu\rho^2}.
\end{equation}
It is seen that the energy levels are shifted by the amount
$\hbar^2s^2/2\mu\rho^2$ which is identical to that predicted by Eq.(22),
so the hydrogen atom in the compactified Kaluza-Klein space $R^1\times S^1$
has the same energy spectrum as that of a hydrogen atom in a one-dimensional
Euclidean space. However, if the length of the compact manifold $S^1$ is
measurable then the energy levels would be different, because in such a
situation the condition $\rho\ll z$ could not be imposed. In that case it
would be possible to detect the difference by a measurement of the frequency
spectrum. Perturbative corrections
to the energy spectrum can be calculated using the generating function for
Laguerre polynomials. In this case the generating function is
\cite{Sned,Brans}
\begin{equation}
-t\exp\left(-\frac{ut}{1-t}\right)=(1-t)^2\sum_{n=1}^\infty
\frac{L_n^1(u)}{n!}t^n.
\end{equation}
The first order correction, $\Delta E_n$, using the term
$e^2\rho^2\phi^2/(2r^3)$ as a perturbation, is given by
\begin{eqnarray}
\Delta E_n&=&\int^{2\pi}_0\int^\infty_{z>0}\psi^*_{ns}(z,\phi)\left( 
\frac{e^2\rho^2}{2z^3}\right)\psi_{ns}(z,\phi)dzd\phi\nonumber\\
&=&\frac{\mu^2e^6\rho^2}{\hbar^4}\frac{(n-1)!}{n^3(n!)^3}
\int^\infty_{u>0}e^{-u}u^{-1}\left[L_n^1(u)\right]^2du\nonumber\\
&=&\frac{\mu^2e^6\rho^2\phi^2}{\hbar^4n^2}Ei(u)
\end{eqnarray}
where $Ei(u)$ is the exponential-integral defined by
\begin{equation}
Ei(u)=\int_{u>0}^\infty\frac{e^{-y}}{y}dy.
\end{equation}
This result shows that perturbative corrections to the unperturbed energy
spectrum can only be carried out for $z>0$. If the hydrogen atom in this
case is measurable only for $z\gg \rho$, then the correction term is
negligible. Higher order perturbative terms can also be calculated and in
general they depend on the quantum number $n$ and integrals of the form
$\int^\infty_{u>0} e^{-y}y^{-2k-1}dy$, where $k=1,2,...$.

\subsection{Quantum mechanics in 3-dimensional space $R^2\times S^1$}

The time-independent Schr\"{o}dinger equation for a free particle in a
three-dimensional compactified Kaluza-Klein space $R^2\times S^1$ can be
written in the form
\begin{equation}
-\frac{\hbar^2}{2\mu}\left(\nabla^2+\frac{1}{\rho^2}\frac{\partial^2}
{\partial \Omega^2}\right)\psi = E\psi,
\end{equation}
where $\rho$ is the radius of the compact circle $S^1$ parametrised by the
angle $\Omega$, and $\nabla^2$ is the Laplacian in two-dimensional Euclidean 
space. If the two-dimensional wavefunction $\psi$ is written in the
form $\psi=\omega(\Omega)\Phi(\phi){\cal R}(r)$, where $(r,\phi)$ are
the polar coordinates in the space R${}^2$, then the above Schr\"{o}dinger
equation reduces to the system of equations
\begin{eqnarray}
\frac{d^2\omega}{d\Omega^2}+s^2\omega&=&0,\\
\frac{d^2\Phi}{d\phi^2}+m^2\Phi&=&0,\\
\frac{d^2{\cal R}}{dr^2}+\frac{1}{r}\frac{d{\cal R}}{dr} -\frac{m^2}{r^2}
{\cal R}+\frac{2\mu}{\hbar^2}\left(E-\frac{\hbar^2s^2}{2\mu\rho^2}
\right){\cal R}&=&0.
\end{eqnarray}
It is seen that a free particle moving in a three-dimensional compactified
Kaluza-Klein space $R^2\times S^1$ posseses an angular momentum
$s$ associated with the third compactified dimension, in addition to the
angular momentum associated with two-dimensional Euclidean space.
An important feature of this extra angular momentum is that it can take on
half-integer values, because the configuration space is multiply connected
and so multivalued wavefunctions are allowed. On the other hand, although
the solution for the function $\Phi$ is of the form 
$\Phi(\phi)=\exp(im\phi)$, the angular momentum $m$ can only
adopt integer values, since in this case the quantity $m$ is associated with
the simply connected Euclidean space R${}^2$. To reiterate, the quantum
dynamics of an electron in a hydrogen atom whose configuration space is
$(R\setminus\{0\})\times S^1$, allows for the angular momentum $m$ to have
half-integer values. The Schr\"{o}dinger equation
for the stationary states of an hydrogen atom in a three-dimensional
compactified Kaluza-Klein space, $R^2\times S^1$, is
\begin{equation}
-\frac{\hbar^2}{2\mu}\left(\nabla^2+\frac{1}{\rho^2}\frac{\partial^2}
{\partial \Omega^2}\right)\psi - \frac{e^2}{r_3}\psi= E\psi
\end{equation}
where $r_3=\sqrt{r^2+\rho^2\Omega^2}$, with $r=\sqrt{x^2+y^2}$. As in
the case of a hydrogen atom in two-dimensional compactified Kaluza-Klein
space, $R^1\times S^1$, the condition $\rho\ll r$ can be imposed, since the
size of the compact space is assumed to be small. The potential can be
expanded as a binormial series and and terms (other than $e^2/r$) may be
regarded as a perturbation. The energy corrections are calculated using the
Laguerre polynomials. Since the calculation is similar to that carried out
in the next section, we postpone a discussion until later. The unperturbed
Schr\"{o}dinger equation for the hydrogen atom reduces to the
system of differential equations
\begin{eqnarray}
\frac{d^2\omega}{d\Omega^2}+s^2\omega&=&0,\\
\frac{d^2\Phi}{d\phi^2}+m^2\Phi&=&0,\\
\frac{d^2{\cal R}}{dr^2}+\frac{1}{r}\frac{d{\cal R}}{dr} -\frac{m^2}{r^2}
{\cal R}+\frac{2\mu}{\hbar^2}\left(\frac{e^2}{r}+E-\frac{\hbar^2s^2}
{2\mu\rho^2}\right){\cal R}&=&0.
\end{eqnarray}
The quantities $s$ and $m$ can take half-integer values, since both
are associated with multiply connected spaces. The
fundamental homotopy group of the space $R^2\setminus \{0\}$ is isomorphic
to $\pi_1(S^1)$, and the fundamental homotopy group of the space
$(R^2\setminus \{0\})\times S^1$ is isomorphic to the fundamental
homotopy group of the space $S^1\times S^1$, which is just the
two-dimensional torus $T^2$, i.e.,  $\pi_1((R^2\setminus \{0\})\times
S^1) \cong \pi_1(S^1\times S^1) \cong \pi_1(S^1)\oplus\pi_1(S^1) \cong
Z\oplus Z$. Therefore, the fundamental group $\pi_1((R^2\setminus\{0\})
\times S^1)$ has a unitary representation
$a: \pi_1((R^2\setminus\{0\}\times S^1)\rightarrow U(1)\times U(1)$ defined
by the character $a(n,m)=\exp(in\pi)\exp(im\pi)$, where $n,m\in Z$. It is
seen that the use of multivalued wavefunctions in this case is
permitted, since the multiple-valued wavefunctions are determined
by the characters of the fundamental homotopy group of the configuration
space.

For a comparison, let us consider quantum mechanics constructed on a
multiply connected configuration space whose fundamental homotopy group is
nonabelian; such an example is that of
the planar hydrogen molecular ion $H_2^+$, with the assumption that
the electron of the system can not penetrate either nucleus. The 
general multiply connected configuration space of this kind has the form
$R^2\setminus \{x_1,...,x_n\}$, where $x_1$,..., $x_n$ are $n$
distinct points in the plane $R^2$. The fundamental homotopy group
$\pi_1$ of the space $R^2\setminus \{x_1,...,x_n\}$ is an infinite
nonabelian group for $n\geq 2$. This is a free group of $n$ generators
which can be constructed by the homotopy classes $[\gamma_i]$ of closed
curves $\gamma_i$ each of which encloses the corresponding point $x_i$ but
none of the remaining points. However, the generators are not determined
uniquely, and when specified, they give rise to a representation of the
group. Hence, the fundamental homotopy group of the configuration space of
the hydrogen molecular ion $H_2^+$ has two generators which can be
identified with two independent loops. It is known that all higher homotopy
groups of $R^2\setminus \{x_1,...,x_n\}$ vanish \cite{Sing,Hilton}.
However, in the case where the atomic electron of the hydrogen molecular
ion, $H_2^+$, is not constrained to the two-dimensional plane
$R^2\setminus\{x_1,...,x_n\}$, all loops are contractible; in other
words, the configuration space of the system is simply connected. In this
case, the wave functions that describe the quantum electronic motion must
be single-valued, and as a consequence, the angular momentum takes on integer
values. Let the origin of the polar coordinates be at the midpoint of two
nuclei which are separated by a distance $d$, then the Schr\"{o}dinger
equation for the stationary states of the electronic motion is written as
\begin{equation}
-\frac{\hbar^2}{2\mu}\nabla^2\psi - \left(\frac{e^2}{r_1}
+ \frac{e^2}{r_2} -\frac{e^2}{d}\right)\psi = E\psi,
\end{equation}
where
$\bf r_1$ and $\bf r_2$ are the position vectors of the electron with
respect to the two protons of the molecule. Using elliptic coordinates
$(\xi,\eta,\psi)$, where $\phi$ is the azimuthal angle with $z$-axis being
the line joining the two protons, $\xi=(r_1+r_2)/d$ and $\eta=(r_1-r_2)/d$,
the Laplacian operator expressed in terms of these coordinates takes the
form \cite{Land,Brans}
\begin{equation}
\nabla^2=\frac{4}{d^2(\xi^2-\eta^2)}\left\{\frac{\partial}{\partial \xi}
(\xi^2-1)\frac{\partial}{\partial\xi} + \frac{\partial} {\partial\eta} (1-
\eta^2)\frac{\partial}{\partial \eta}  + \frac{\xi^2- \eta^2}{(\xi^2-1)(1-
\eta^2)} \frac{\partial^2}{\partial \phi^2}\right\}.
\end{equation}
When the wave function is written as a product
$\psi=\Phi(\phi)F(\xi)G(\eta)$, the Schr\"{o}dinger equation reduces to the
system of equations, in atomic units $\mu=e=\hbar=a_0=1$,
\begin{eqnarray}
\frac{d^2\Phi}{d\phi^2}+m^2\Phi&=&0\\
\frac{d}{d\xi} (\xi^2-1)\frac{dF}{d\xi} + \left(\frac{d^2}{2}
\left(E-\frac{1}{d}\right)\xi^2 + 2d\xi - \frac{m^2}{\xi^2-1} + \lambda
\right)F(\xi)&=&0\\
\frac{d}{d\eta}(1-\eta^2)\frac{dG}{d\eta} -\left(\frac{d^2}{2}
\left(E-\frac{1}{d}\right)\eta^2 + \frac{m^2}{1-\eta^2}+
\lambda\right)F(\xi)&=&0,
\end{eqnarray}
where $m$ and $\lambda$ are separation
constants. The solution to the Eq.(42) is of the form $\Phi=\exp(im\phi)$ 
and in this case $m$ must take integer values because the fundamental group 
of the configuration space vanishes.

\subsection{Quantum mechanics in 4-dimensional space $R^3\times S^1$}

The time-independent Schr\"{o}dinger equation for a free particle in a
four-dimensional compactified Kaluza-Klein space $R^3\times S^1$ can be
written in the form
\begin{equation}
-\frac{\hbar^2}{2\mu}\left(\nabla^2+\frac{1}{\rho^2}\frac{\partial^2}
{\partial \Omega^2}\right)\psi_4 = E\psi_4,
\end{equation}
where $\rho$ is the radius of the compact circle $S^1$ parametrised by the
angle $\Omega$, and $\nabla^2$ is the Laplacian in three-dimensional
Euclidean space. If the four-dimensional wavefunction $\psi_4$ is written
in the
form $\psi_4=\omega(\Omega)\psi(r,\theta,\phi)$, where $(r,\theta,\phi)$ are 
the three-dimensional spherical coordinates, then the above Schr\"{o}dinger
equation reduces to the system of equations
\begin{eqnarray}
\frac{d^2\omega}{d\Omega^2}+s^2\omega&=&0,\\
\frac{\hbar^2}{2\mu}\nabla^2\psi+k^2\psi&=&0,
\end{eqnarray}
where $k$ is defined by $E=\hbar^2k^2/2\mu-\hbar^2s^2/2\mu\rho^2$.
As in the case of a free particle in three-dimensional compactified
Kaluza-Klein space, $R^2\times S^1$, the Schr\"{o}dinger equation 
gives rise to an angular momentum $s$ which can take on
half-integer values. The energy spectrum is also shifted by an amount
$\hbar^2s^2/2\mu\rho^2$. Therefore, free particle eigenfunctions in
four-dimensional compactified Kaluza-Klein space $R^3\times S^1$ can be
classified by the continuous energy eigenvalues $E$ and three discrete
indices $s, l$ and $m$, where the quantum numbers $l$ and $m$ result
from the three-dimensional free-particle solutions $\psi_{Elm}({\bf r}) =
j_{El}(kr)Y_{lm}(\theta,\phi)$. Both of the quantum
numbers $l$ and $m$ are integers since they are associated only with the
simply connected Euclidean space R${}^3$.

The Schr\"{o}dinger equation for the stationary states of an hydrogen atom
in a four-dimensional compactified Kaluza-Klein space, $R^3\times S^1$,
can be written in the form
\begin{equation}
-\frac{\hbar^2}{2\mu}\left(\nabla^2+\frac{1}{\rho^2}\frac{\partial^2}
{\partial \Omega^2}\right)\psi_4 - \frac{e^2}{r_4}\psi_4 = E\psi_4,
\end{equation}
where $r_4=\sqrt{r^2+\rho^2\Omega^2}$ with $r=\sqrt{x^2+y^2+z^2}$. As in
the case of a hydrogen atom in two-dimensional compactified Kaluza-Klein
space, the condition $\rho\ll r$ is imposed, since the size of the
compact space is assumed to be small.
The potential is expanded in a binomial series as
\begin{equation}
V=\frac{e^2}{r}\left[1-\frac{1}{2}\left(\frac{\rho\Omega}{r}\right)^2 +
\frac{3}{8}\left(\frac{\rho\Omega}{r}\right)^4 -
\frac{5}{16}\left(\frac{\rho\Omega}{r}\right)^6 + ... \right].
\end{equation}
If the terms that contain the quantity $\rho\Omega$  are treated as a
perturbation, then the unperturbed Schr\"{o}dinger equation for the hydrogen
atom reduces to the system of differential equations
\begin{eqnarray}
\frac{d^2\omega}{d\Omega^2}+s^2\omega&=&0,\\
\frac{\hbar^2}{2\mu}\left(\nabla^2+\frac{e^2}{r}+k^2\right)\psi&=&0.
\end{eqnarray}
The behaviour of an hydrogen atom in four-dimensional compactified
Kaluza-Klein space is therefore identical to that of a hydrogen atom in
three-dimensional Euclidean space, since the ground state energy
$\hbar^2s^2/2\mu\rho^2$ is unobservable. However, unlike the situation in
three-dimensional Euclidean space, the Schr\"{o}dinger equation in
four-dimensional compactified Kaluza-Klein space gives rise to an
angular momentum that can take on half-integer values, which hints at a
possible topological origin of the spin of the electron. It should be
emphasised again that the half-integer values of the angular momentum $s$
are possible because the background space $R^3\times S^1$ is multiply
connected.

If the terms in the binomial series of the potential, that contain the
quantity $\rho\Omega$, are treated perturbatively then their correction to
the energy spectrum can be calculated from
\begin{eqnarray}
\left<r^s\right>_{nlm}&=&\int\psi^*_{nlm}({\bf r})r^s\psi_{nlm}({\bf r})
d{\bf r}\nonumber\\
&=&\int^\infty_0r^{2+s}|R_{nl}(r)|^2dr,
\end{eqnarray}
where the radial wavefunctions $R_{nl}(r)$ are defined by
\begin{equation}
R_{nl}(r)=-\left\{\left(\frac{2}{na_0}\right)^3\frac{(n-l-1)!}{2n [(n+
l)!]^3}\right\}^{1/2}e^{-\rho/2}\rho^l L^{2l+1}_{n+l}(\rho)
\end{equation}
with $L^{2l+1}_{n+l}$ being the associated Laguerre polynomial and
\begin{equation}
\rho=\frac{2}{na_\mu}r, \ \ \ \ \ a_0=\frac{4\pi\epsilon_0\hbar^2}{\mu
e^2}.
\end{equation}
Here $a_0$ denotes the Bohr radius and $\mu$ is the reduced mass of the
system.

For the case $s=-1$ and $s=-2$, using a generating function for the
associated Laguerre polynomials $L_q^p(\rho)$
\begin{eqnarray}
G_p(\rho,s)&=&\frac{(-s)^p\exp[-\rho s/(1-s)]}{(1-s)^{p+1}}\nonumber\\
&=&\sum^\infty_{q=p}\frac{L_q^p(\rho)}{q!}s^q
\end{eqnarray}
it can be shown that \cite{Sned,Brans}
\begin{eqnarray}
\left<\frac{1}{r}\right>_{nlm}&=&\frac{1}{a_0 n^2}\\
\left<\frac{1}{r^2}\right>_{nlm}&=&\frac{1}{a_0^2n^3(l+1/2)}
\end{eqnarray}
However for $s\leq -3$ the following recursion relation can be used
\cite{Mess}
\begin{equation}
\frac{s+1}{n^2}<r^s>-(2s+1)a_0<r^{s-1}>+\frac{s}{4}[(2l+1)^2-s^2]a_0^2
<r^{s-2}>=0,
\end{equation}
with the condition $s>-2l-1$. In this case it is found, for example, that
\begin{equation}
\left<\frac{e^2\rho^2\Omega^2}{2}\frac{1}{r^3}\right>_{nlm}=\frac{e^2\rho^2
\Omega^2}{2a_0^3n^3l(l+1/2)(l+1)}
\end{equation}
Because these corrections involve the quantity $\rho$, the above results
show that the corrections only become significant when the compact
dimension is measurable. When the size of the compact space is unmeasurable
all energy corrections can be ignored and the hydrogen atom in this case
behaves like a hydrogen atom in ordinary three-dimensional Euclidean space.

\section{Conclusion}

We have analysed quantum mechanics in multiply connected spaces emphasising
the role which topology may play in determining the nature of a quantum
observable. Based on the fact that the configuration
space of physical systems, such as the planar hydrogen atom or the planar
molecular ion $H^+_2$, is multiply connected, it has been argued
that the angular momentum of those physical systems can adopt half-integer
values, which hints at a topological origin of the spin of the electron. In
order to incorporate topological constraints into the dynamics of the
electron in three-dimensional space we have considered nonrelativistic
quantum mechanics in a compactified Kaluza-Klein space consisting of the
direct product of an $(N-1)$-dimensional Euclidean space with the compact
circle S${}^1$. The configuration space in this form, whose universal
covering space is the simply connected Euclidean space R${}^N$, allows
the introduction of the intrinsic spin into the nonrelativistic
Schr\"{o}dinger wave equation in a simple manner.
Furthermore, the hydrogen atom in the background space of compactified
Kaluza-Klein space has an energy spectrum that differs from that in the
background Euclidean space. If the size of the compactified space
is unmeasurable then the two energy spectra are not distinguishable, which
is shown by carrying out a simple perturbation calculation. However, this
result does not exclude the possibility that spin is a
manifestation of the compact dimension of the spacetime manifold.

\section*{Acknowledgements}
We would like to thank the referees for their constructive comments.
We would also like to thank C R Hagen for a private communication on the
two-dimensional hydrogen atom and references related to this problem.
One us (VBH) acknowledges the financial support of an APA Research Award.

\end{document}